\documentclass[3p,twocolumn]{elsarticle}
\usepackage{lineno,hyperref}
\usepackage{amssymb}
\usepackage{graphicx}
\usepackage{amsmath}
\usepackage{color}
\usepackage{float}
\usepackage{mathrsfs}
\usepackage{indentfirst}
\usepackage{textcomp}
\usepackage{comment}
\usepackage{mathtools}

\newcommand{\ket}[1]{\left| #1 \right>} 
\newcommand{\bra}[1]{\left< #1 \right|} 
\newcommand{\braket}[2]{\left< #1\vphantom{#2}\vert#2\vphantom{#1} \right>} 

\newcommand{\COMMENTED}[1]{}

\biboptions{sort&compress}

\begin{document}

\title{Accurate computations of Rashba spin-orbit coupling in interacting systems: \\
from the Fermi gas to real materials}

\author[wm]{Peter Rosenberg}
\author[wm,ccp]{Hao Shi}
\author[wm]{Shiwei Zhang}

\address[wm]{Department of Physics, The College of William and Mary, Williamsburg, Virginia 23187}
\address[ccp]{Center for Computational Quantum Physics, The Flatiron Institute, 162 5th Avenue, New York, NY 10010}

\begin{abstract}
We describe the treatment of Rashba spin-orbit coupling (SOC) in interacting many-fermion systems within the 
auxiliary-field quantum Monte Carlo framework, and present a set of illustrative results. These include numerically exact 
calculations on the ground-state properties of the spin-balanced, attractive two-dimensional Fermi gas,
as well as a study of a tight-binding Hamiltonian with repulsive interaction.
These systems are formally connected via the Hubbard Hamiltonian with SOC,
but cover different physics ranging from superfluidity and triplet pairing to SOC in 
real materials in the presence of strong interactions in localized orbitals. 
We carry out detailed benchmark studies of the method  in the
latter case when an approximation is needed to control the sign problem for repulsive Coulomb interactions.
The methods presented here provide an approach for predictive computations 
in materials to study the interplay of SOC and strong correlation.
\end{abstract}

\maketitle

\section{Introduction}

Spin-orbit coupling lies at the heart of a tremendous variety of physical phenomena, 
in contexts ranging from semiconductors and metals to 2D heterostructures and even ultracold atoms. 
The field has garnered renewed interest recently due to the central role that SOC plays in many exotic 
topological phases, including the quantum spin Hall effect, topological insulators and superconductors, 
and Majorana fermions \cite{reviewRashbaMaterials,reviewRashba_nat,reviewRashba_rep_prog_phys}.  
These developments underpin modern progress in spintronics \cite{Koo1515,reviewSpintronics}, have opened the field of spin-orbitronics, 
and carry important implications for quantum computation and information \cite{reviewQuantumSpintronics}.

Recent progress in the understanding and manipulation of SOC in real materials has been greatly
complemented by rapid and remarkable progress in the field of cold atom physics.
Two notable achievements are the ability to tune interaction strengths in ultracold gases using Feshbach 
resonances, and the ability to load ultracold gases into optical lattice potentials \cite{reviewBloch,
reviewLewenstein}. These techniques have ushered in a new era of quantum engineering and 
simulation, and provide an ideal platform to study many-body physics \cite{2DFG_expt,
2DFG_AFQMC,2DFG_DMC,2DFG_FINITET_PRL}. A wide variety of lattice models have already been 
experimentally realized, with even more likely in the near future. These clean systems, with a high degree of 
experimental control, offer a novel means of exploring exotic states of matter. Many of the difficulties encountered in 
real materials, including fixed structural and electromagnetic properties, and disorder, are surmounted in
ultracold atom experiments, which can be freely tuned to probe broad parameter regimes. 
The third major step was the development of methods to generate ``synthetic SOC''
in ultracold atoms \cite{artificialgaugefieldJaksch, PhysRevLett.95.010403,PhysRevLett.108.225303,
PhysRevLett.107.255301,PhysRevLett.111.185302, GerbierArtifcialGaugeFields},
which has created new paths towards the realization of many of the exotic states long sought after in real materials, 
such as high-$T_c$ superconductivity and topological phases, including quantum spin Hall states and Majorana fermions. 
These techniques have enabled experimental emulations of a wide variety of fundamental models, including several 
with Rashba, Dresselhaus, or equal Rashba-Dresselhaus SOC \cite{SOC_BEC_2009,SOC_BEC_2011,SOC_DFG_CHEUK,SOC_DFG_WANG,
RASHBASOC2D_EXPT}.    

This expanding experimental horizon has generated tremendous opportunities for theoretical and numerical 
approaches to many-body systems. 
The delicate nature of many of the exotic states demands high-accuracy
theoretical and numerical input. 
Careful characterization of these phases, and their behavior in the presence of strong interactions, will 
provide crucial benchmarks for experiments as well as essential guidelines for the design and fabrication 
of novel devices \cite{reviewSpinHallDevices}. 
These systems are typically quite challenging from a theoretical or numerical 
perspective.
$\emph{Ab initio}$ treatment of strongly-correlated real 
materials remains an outstanding challenge in many-body physics. 
However, ultracold atom experiments, which can be measured to high accuracy, offer a unique opportunity to 
test and calibrate our theoretical and computational methods. 

The auxiliary-field quantum Monte Carlo (AFQMC) method is a general computational approach for interacting
many-fermion systems.
AFQMC has been applied to a variety of electronic models and materials without SOC, including Hubbard-like models 
\cite{MingpuHubbardBenchmark, Hao-symmetry},
molecular systems \cite{AFQMC_backprop_molecules,HydrogenBenchmark,Mo2_AFQMC}, and solids 
\cite{AFQMC_downfolding}. In ultracold atoms, 
a number of recent successes have highlighted the accuracy and capability of AFQMC,
including studies of the attractive spin-balanced Fermi gas \cite{2DFG_AFQMC, Ettore-gaps} and fully treating interaction and Rashba SOC on the same footing \cite{2DFG_SOC_AFQMC,2DFG_RASHBASOC_OPLATT_AFQMC}. The Hamiltonian for the spin-balanced system with contact interaction preserves time-reversal symmetry, 
thus guaranteeing that simulations are free of the sign problem \cite{nosignproblem}, so the results obtained
are numerically exact. This is a particularly notable achievement given the emergence of quantum gas 
microscopes with single-site resolution \cite{PhysRevLett.116.235301, Greif953, Cheuk2015, Haller2015, 
Parsons2015}, which provide a precise calibration of the technique via direct comparison with a clean, tunable
system. 

A natural next step is to treat electronic Hamiltonians with SOC, which is the topic of focus  in the 
present paper. 
We present an illustrative set of results on the attractive 2D Fermi gas with Rashba SOC,
as well as the first study of Rashba SOC in a real material context using AFQMC. 
From the ultracold atom Hamiltonians we have treated, an 
immediate connection to real materials can be made by considering systems with repulsive interactions. 
We demonstrate
the capabilities of the AFQMC method by considering a repulsive Hubbard model, which captures much
of the underlying physics of real materials.
In this way, the two types of systems we study here, 
the real material and ultracold atoms,  
are described by a similar Hamiltonian, namely the Hubbard model, with
opposite sign interactions. This Hamiltonian is of fundamental importance in condensed matter 
and many-body physics. It can be engineered and emulated  by optical lattice experiments  \cite{2DHubbard_Hulet}, with 
the sign of the interaction controlled using Feshbach
resonances. This would permit a direct comparison of experimental 
results with our calculations, and consequently a better understanding of 
the physics of both systems, as well as an additional calibration of the AFQMC method. 

With the repulsive interaction in the Hubbard model, the sign problem reemerges in AFQMC, but as we will show, this can be systematically controlled by applying a constraint \cite{CPMC}. The accuracy achieved is comparable to what has been systematically seen in real materials 
without SOC. 
Since the AFQMC method provides access to a wide range of observables
to probe the charge, spin,
pairing, and transport properties,
we expect this development to lead to significant applications in correlated materials with strong SOC. 

We organize the remainder of the paper as follows. In section \ref{sec:Hamiltonian} we 
introduce the general Hamiltonian used to describe systems with Rashba SOC, and
establish its connection to both real materials and the Fermi gas. Section \ref{sec:methods}
outlines the AFQMC method, highlighting several of the recent advances incorporated to treat SOC.
Two different sampling approaches are discussed, for cases without and with the sign problem, respectively.
Section \ref{sec:results} provides a demonstration of the technique, first in the context of
cold atoms and then on a repulsive Hubbard model, which is closely connected to the
behavior of real materials. Finally, section \ref{sec:summary} summarizes and offers an outlook on open
questions and promising directions in this burgeoning field. 

\section{Hamiltonian}
\label{sec:Hamiltonian}
We begin with a  form of the  SOC Hamiltonian on a lattice, 
\begin{align}
\hat{H}=&\sum_{\mathbf{k}\sigma}\varepsilon_{\mathbf{k}}c^\dagger_{\mathbf{k}\sigma} c_{\mathbf{k}\sigma} 
+ \sum_{\mathbf{k}} \left(\mathcal{L}_\mathbf{k} c^\dagger_{\mathbf{k}\downarrow} c_{\mathbf{k}\uparrow} 
+ \textrm{h.c.}\right) \notag
\\& +\sum_\mathbf{i} U n_{\mathbf{i}\uparrow}n_{\mathbf{i}\downarrow}\,,
\label{eq:Hamiltonian}
\end{align}
where $c^\dagger_{\mathbf{k}\sigma}$ creates a spin-$\sigma$ ($=\uparrow$ or $\downarrow$) particle with 
momentum $\mathbf{k}$, and $n_{i\sigma}=c^\dagger_{i\sigma} c_{i\sigma}$ denotes the density operator in 
real-space on site $i$ with spin-$\sigma$. 
Although we have used the Rashba form of SOC,
 our discussions are general for other types.
For two-dimensional lattice systems the dispersion and SOC terms are $\varepsilon_\mathbf{k} = -2t(\cos k_x+\cos k_y)$ and
$\mathcal{L}_\mathbf{k} = 2\lambda(\sin k_y-i\sin k_x)$. The parameter $t$, set to unity throughout this work, 
determines the strength of nearest-neighbor hopping, the parameter $\lambda$ controls the strength of SOC,
and the parameter $U$ determines the strength of the on-site interaction.
For the Fermi gas, which represents the low-density limit of the lattice model,
$\varepsilon_\mathbf{k} = \mathbf{k}^2$ and $\mathcal{L}_\mathbf{k} = \lambda(k_y-ik_x)$.
In both cases, a natural description of the
system in terms of helicity bands, $\varepsilon^\pm_\mathbf{k}=\varepsilon_\mathbf{k}\pm\vert 
\mathcal{L}_\mathbf{k} \vert$,
is obtained by diagonalizing the non-interacting Hamiltonian.

The Fermi gas Hamiltonian is defined in the dilute limit of Eq.~(\ref{eq:Hamiltonian}), and
the interaction $U/t$ is negative: $U/t<0$. 
In order to directly compare with 
experiments,
the parameters $t$ and $U$ are fully specified, as we further discuss in Sec.~\ref{ssec:Fermigas}.
The physics is dictated by the quantity $k_F a$, 
where the Fermi wave-vector $k_F$ measures the inverse of the average inter-particle spacing while $a$ is the scattering length. As $k_F a>0$ decreases in 2D, the system undergoes a BCS-BEC crossover. Our study will
examine the interplay between this effect and SOC, which induces triplet pairing.

For the second application, namely as a model for electronic systems, we will consider $U/t>0$, and 
consider a range of values for $U/t$ from $0$ to about $8$, which represents strong local interactions
as is typical in models for cuprates. Most of our calculations will be at intermediate interaction $U/t=4$,
and we examine different regimes of SOC strength by varying $\lambda$. Our study here is mostly 
for testing the algorithm. In most cases we focus on small lattice sizes for which we can obtain 
exact results from exact diagonalization.
With this description of the model for both cases, we can now outline its treatment within the AFQMC framework.

\section{Methods}
\label{sec:methods}

\subsection{Preliminaries}
\label{ssec:background}

We will first provide an overview of the AFQMC method \cite{Lecture-notes,BSS,Koonin},
and then discuss some of the extensions necessary to treat systems with SOC. In general, 
ground-state QMC methods rely on imaginary-time projection to obtain the many-body ground-state 
$\ket{\Psi_0}$ of a given hamiltonian $\hat{H}$ from a starting trial wavefunction 
$\ket{\Psi_T}$. This projection proceeds according to,
\begin{equation}
\ket{\Psi_0} \propto \underset{\beta\rightarrow\infty}{\lim} e^{-\beta \hat{H}}\ket{\Psi_T},
\label{eq:proj_lim}
\end{equation}
provided $\braket{\Psi_T}{\Psi_0}\neq 0$ (i.e. the trial wavefunction
cannot be orthogonal to the many-body ground-state).

In order to carry out this projection numerically, we first discretize the imaginary
time interval into $m = \beta/\Delta\tau$ time slices,
\begin{equation}
e^{-\beta\hat{H}}=\left(e^{-\Delta\tau\hat{H}}\right)^m,
\end{equation}
so that the limit in (\ref{eq:proj_lim}) can be obtained iteratively via,
\begin{equation}
\vert\Psi^{(n+1)}\rangle=e^{-\Delta\tau\hat{H}}\vert\Psi^{(n)}\rangle,
\label{eq:proj_iter}
\end{equation}
with $\ket{\Psi^{(0)}}=\ket{\Psi_T}$.

To proceed, we must rewrite the many-body propagator $e^{-\Delta\tau\hat{H}}$  
in a single-particle form. This is accomplished by applying the Trotter-Suzuki 
breakup,
\begin{equation}
\left(e^{-\Delta\tau\hat{H}}\right)^m = 
\left(e^{-\Delta\tau\hat{K}/2}e^{-\Delta\tau\hat{V}}e^{-\Delta\tau\hat{K}/2}\right)^m
+ \mathcal{O}(\Delta\tau^2),
\end{equation}
where $\hat{K}$ contains the one-body terms and $\hat{V}$ the two-body terms of the Hamiltonian
in (\ref{eq:Hamiltonian}).
This step is followed by a suitable Hubbard-Stratonovich (HS) transformation 
\cite{continuous-HS-transformation}. We list below four varieties of discrete HS transformation 
commonly used in AFQMC simulations:
The charge decomposition \cite{HS_transform_discrete} is written,
\begin{equation}
e^{\Delta\tau Un_{i\uparrow}n_{i\downarrow}}=
\frac{1}{2}\sum_{x_i=\pm1}e^{(\gamma x_i-\Delta\tau U/2)(n_{i\uparrow}+n_{i\downarrow}-1)},
\end{equation}
with $\gamma$ determined according to $\cosh(\gamma)=\exp(-\Delta\tau\,U/2)$;
The spin decomposition in the $z$- or $x$-direction has the form,
\begin{equation}
\label{eq:HS-Sz-Sx}
e^{\Delta\tau Un_{i\uparrow}n_{i\downarrow}}=
\frac{1}{2}\sum_{x_i=\pm1}e^{2\gamma x_i S_i^{z, x} -\Delta\tau U(n_{i\uparrow}+n_{i\downarrow})/2},
\end{equation}
and the spin decomposition in the $y$ direction is,
\begin{equation}
\label{eq:HS-Sy}
e^{\Delta\tau Un_{i\uparrow}n_{i\downarrow}}=
\frac{1}{2}\sum_{x_i=\pm1}e^{2 \mathrm{i} \gamma x_i S_i^{y}-\Delta\tau U(n_{i\uparrow}+n_{i\downarrow})/2}.
\end{equation}
A decomposition in the $x$-$z$ plane can also be obtained by using a linear combination of $S_i^{z}$ and $S_i^{x}$.
In Eqs~(\ref{eq:HS-Sz-Sx}) and (\ref{eq:HS-Sy}),  
$\cos(\gamma)=\exp(\Delta\tau\,U/2)$, and the spin operators are defined as:\\
$S_i^{z}=(c^\dagger_{i\uparrow}c_{i\uparrow}-c^\dagger_{i\downarrow}c_{i\downarrow})/2$;
$S_i^{x}=(c^\dagger_{i\uparrow}c_{i\downarrow}+c^\dagger_{i\uparrow}c_{i\downarrow})/2$;
$S_i^{y}=(c^\dagger_{i\uparrow}c_{i\downarrow}-c^\dagger_{i\uparrow}c_{i\downarrow})/2i$.

This procedure yields the following form for the propagator,
\begin{equation}
e^{-\Delta\tau\hat{H}}=\int d\mathbf{x}\,p(\mathbf{x})\hat{B}(\mathbf{x}),
\label{eq:prop_int}
\end{equation}
where $\mathbf{x}=\{x_1,x_2,\hdots,x_{N_S}\}$ is a
set of auxiliary fields at a given time slice, with dimension $N_S$ equal to the
size of the single-particle basis, which in the case of lattice systems is typically the number
of lattice sites.
Using the charge decomposition as an example, the probability density function $p(\mathbf{x})$ is 
uniform, and, 
\begin{equation}
\hat{B}(\mathbf{x}) \equiv e^{-\Delta\tau\hat{K}/2}
 \prod_i \hat{b}_i(x_i)
 e^{-\Delta\tau\hat{K}/2},
\end{equation}
with $\hat{b}_i(x_i) \equiv  \exp\left[{(\gamma x_i-\Delta\tau U/2)(n_{i\uparrow}+n_{i\downarrow}-1)}\right]$.
The many-body propagator is now composed of single particle operators
with the fermions in external auxiliary fields. The integration over auxiliary field configurations recovers the two-body
interactions.

\subsection{Metropolis sampling of paths in AF space}
\label{ssec:Metrop}

Ground-state observables are calculated according to,
\begin{equation}
\langle \hat{A} \rangle = \frac{\bra{\Psi_T}e^{-\beta\hat{H}/2}\hat{A}e^{-\beta\hat{H}/2}\ket{\Psi_T}}
{\bra{\Psi_T}e^{-\beta\hat{H}}\ket{\Psi_T}}.
\label{eq:pi}
\end{equation}
The denominator in (\ref{eq:pi}) is,
\begin{align}
&\int\bra{\Psi_T} \prod_{\ell=1}^m d\mathbf{x}^{(\ell)}p(\mathbf{x}^{(\ell)})\hat{B}(\mathbf{x}^{(\ell)})\ket{\Psi_T}\notag\\
\equiv&\int\mathcal{W}(\mathbf{X})d\mathbf{X}\,,
\label{eq:pi_denom}
\end{align}
where,
\begin{equation}
\mathcal{W}(\mathbf{X})=\braket{\Psi_l}{\Psi_r}\prod_{\ell=1}^m p(\mathbf{x}^{(\ell)}),
\label{eq:W}
\end{equation}
and we have introduced the notation,
\begin{align*}
\bra{\Psi_l}&=\bra{\Psi_T}\hat{B}(\mathbf{x}^{(m)})\hat{B}(\mathbf{x}^{(m-1)})\hdots\hat{B}(\mathbf{x}^{(n)})\\
\ket{\Psi_r}&=\hat{B}(\mathbf{x}^{(n-1)})\hat{B}(\mathbf{x}^{(n-2)})\hdots\hat{B}(\mathbf{x}^{(1)})\ket{\Psi_T}.
\end{align*}
In the above, $\mathbf{x}^{(\ell)}$ represents an auxiliary field configuration at time slice $\ell$,
and the collection of auxiliary fields $\mathbf{X}=\{\mathbf{x}^{(1)},\mathbf{x}^{(2)},\hdots,\mathbf{x}^{(m)}\}$
comprises a path in auxiliary field space.

With this shorthand, Eq.~(\ref{eq:pi}) can be written as a path integral over auxiliary fields,
\begin{equation}
\langle \hat{A} \rangle = \frac{\int\mathcal{A}(\mathbf{X})\mathcal{W}(\mathbf{X})d\mathbf{X}}
{\int\mathcal{W}(\mathbf{X})d\mathbf{X}},
\label{eq:pi_shorthand}
\end{equation}
with,
\begin{equation}
\mathcal{A} = \frac{\bra{\Psi_l}\hat{A}\ket{\Psi_r}}{\braket{\Psi_l}{\Psi_r}}.
\end{equation} 
This integral can be evaluated using standard
Monte Carlo techniques, such as the Metropolis algorithm, which
samples auxiliary-fields from $\mathcal{W}(\mathbf{X})$ to obtain a
Monte Carlo estimate of the expectation value in (\ref{eq:pi_shorthand}).
To accelerate the sampling procedure we employ a dynamic force
bias \cite{2DFG_AFQMC, Lecture-notes}, which improves the acceptance ratio 
and consequently the efficiency of the algorithm. In addition, we remove the
infinite variance problem using the bridge link method~\cite{inf_var}.

\subsection{Branching random walks and the constraint}
\label{ssec:CPMC}

In systems that preserve time-reversal symmetry, such as the spin-balanced attractive Hubbard model 
(even in the presence of SOC, as discussed below) or
the half-filled repulsive Hubbard model, the denominator $\mathcal{W}(\mathbf{X})$ remains positive
and the path-integrals in Eq.~(\ref{eq:pi}) can be calculated with the Monte Carlo technique outlined above, which
samples the probability density function via the Metropolis algorithm. In cases where time-reversal symmetry
is broken,  $\mathcal{W}(\mathbf{X})$ is no longer guaranteed to be positive, and the Monte Carlo signal obtained
by this straightforward sampling procedure is lost to sampling noise. This is a manifestation of the 
well-known sign problem \cite{Lecture-notes}. 

In order to treat systems with a sign problem we recast the procedure outlined above as an open-ended random 
walk in Slater-determinant space, which then allows the imposition of a constraint to prevent the decay of the Monte Carlo signal \cite{CPMC}. Returning to 
(\ref{eq:proj_iter}), we have,
\begin{align}
\vert\Psi^{(n+1)}\rangle&=e^{-\Delta\tau\hat{H}}\vert\Psi^{(n)}\rangle\notag\\
&=\int d\mathbf{x}\,p(\mathbf{x})\hat{B}(\mathbf{x})\vert\Psi^{(n)}\rangle. \label{eq:proj_iter_int}
\end{align}
In this formulation the wave function at each step is represented by an ensemble of Slater determinants,
\begin{equation}
\vert\Psi^{(n)}\rangle\propto\sum_k w_k^{(n)}\vert\phi_k^{(n)}\rangle,
\end{equation}
where $w_k^{(n)}$ denotes the weight for the $k$-th walker at time step $n$, and the sum
runs over the entire walker population at that time step.

To eliminate the sign problem we impose a constrained-path (CP) approximation  \cite{CPMC,Lecture-notes}, which
requires at each time step that all walkers maintain positive overlap 
with $\ket{\Psi_T}$,
\begin{equation}
\langle\Psi_T\vert\phi_k^{(n)}\rangle>0.
\label{eq:constraint}
\end{equation}
To implement this constraint within the random walk procedure we define an importance function,
\begin{equation}
O_T(\phi)\equiv \textrm{max}\{\langle\Psi_T\vert\phi\rangle,0\},
\end{equation}
which prevents walkers from acquiring a negative overlap with $\ket{\Psi_T}$.

With the addition of importance sampling we carry out a modified version of the random walk,
\begin{equation}
\vert\tilde{\Psi}^{(n+1)}\rangle = \int d\mathbf{x}\,\tilde{p}(\mathbf{x})\hat{B}(\mathbf{x})\vert\tilde{\Psi}^{(n)}\rangle,
\label{eq:proj_iter_int_mod}
\end{equation}
with modified probability density function,
\begin{equation}
\label{eq:imp-sampl}
\tilde{p}(\mathbf{x})=p(\mathbf{x})\frac{O_T(\phi^{(n+1)})}{O_T(\phi^{(n)})},
\end{equation}
and modified wavefunction,
\begin{equation}
\vert\tilde{\Psi}^{(n)}\rangle\propto\sum_k w_k^{(n)}\vert\phi_k^{(n)}\rangle.
\end{equation}
The true wavefunction is then,
 \begin{equation}
 \vert\Psi^{(n)}\rangle\propto\sum_k w_k^{(n)}\frac{\vert\phi_k^{(n)}\rangle}{O_T(\phi^{(n)})}.
 \end{equation}
The modified projection equation written in (\ref{eq:proj_iter_int_mod}) is 
identical to the version in (\ref{eq:proj_iter_int}). In the modified version auxiliary fields 
are sampled from $\tilde{p}(x)$, which favors determinants with larger overlaps 
with $\vert \Psi_T \rangle$, and vanishes if the overlap is zero, thus enforcing the 
constraint in (\ref{eq:constraint}). After the random walk has equilibrated, the
ground state wavefunction is represented by the distribution of weighted random
walkers. 

It is important to note that the branching random walk approach discussed here is
closely related to the Metropolis sampling discussed in Sec.~\ref{ssec:Metrop}.
 (The dynamic force bias we use in the Metropolis sampling has the same origin and the same 
form as that obtained from the importance sampling of  in Eq.~(\ref{eq:imp-sampl}).)
The only difference is that the Metropolis procedure, while more convenient for 
computing observables, encounters severe ergodicity problems when the constraint has to be 
imposed. In the current approach, 
the total energy is calculated in a manner similar to that in Sec.~\ref{ssec:Metrop}, except $\langle\Psi_l |$ is fixed to be $\langle\Psi_T|$. For 
computing observables that do not commute with the Hamiltonian, we use 
the backpropagation technique \cite{CPMC, AFQMC_bosons, AFQMC_backprop_molecules}.

\subsection{Generalizations to SOC}
\label{ssec:SOC}

Having outlined the standard AFQMC procedures, we can now address the modifications necessary
to treat SOC. In systems without SOC, there is no mixing of different spin sectors, so random walkers
can be factored into $\uparrow$- and $\downarrow$-spin components,
 \begin{equation}
\ket{\phi} = \ket{\phi^\uparrow} \otimes \ket{\phi^\downarrow}. 
\end{equation}
For a system of $N_\sigma$ particles with single particle orbitals of dimension $N_s$,  
the walker $\ket{\phi^\sigma}$ is just a $N_s \times N_\sigma$ Slater determinant, $\Phi^\sigma$, 
with matrix form,
\begin{equation}
\begin{pmatrix}
\varphi^\sigma_{1,1} && \varphi^\sigma_{1,2} && \hdots && \varphi^\sigma_{1,N_\sigma}\\
\varphi^\sigma_{2,1} && \varphi^\sigma_{2,2} && \hdots  && \varphi^\sigma_{2,N_\sigma}\\
\vdots && \vdots && && \vdots \\
\varphi^\sigma_{N_s,1} && \varphi^\sigma_{N_s,2} && \hdots && \varphi^{\sigma}_{N_s,N_\sigma}  
\end{pmatrix}.
\label{eq:phi_sigma}
\end{equation}
The one body propagator used to project these random walkers can also 
be split into $\uparrow$- and $\downarrow$-spin components,
\begin{equation}
\hat{B}(\mathbf{x})=\hat{B}^\uparrow(\mathbf{x})\otimes\hat{B}^\downarrow(\mathbf{x}),
\end{equation}
where $\hat{B}^\uparrow(\mathbf{x})$ contains only spin-$\uparrow$ operators, and
$\hat{B}^\downarrow(\mathbf{x})$ contains only spin-$\downarrow$. 
We set $\mathbb{B}^\sigma$ to be the matrix representation of $\hat{B}^{\sigma}(\mathbf{x})$, which is a $N_s\times N_s$ square matrix. 

The above applies to both the Metropolis and the branching random walk  approaches. In the 
Metropolis approach discussed in Sec.~\ref{ssec:Metrop},
each path preserves the original spin sectors defined by $|\Psi_T\rangle$, and 
both $\langle \Psi_l|$ and $|\Psi_r\rangle$ will take the same form. In the branching random walk 
approach discussed in Sec.~\ref{ssec:CPMC}, each walker will remain in the same form.

When SOC is included the simple factorization into separate spin components is no longer possible.
Instead, each random walker must be in generalized Hartree-Fock (GHF) form consisting of spin-orbitals, which has
a $2N_s\times N$ matrix representation $\Phi$, 
\begin{equation}
\begin{pmatrix}
\varphi_{1,1} && \varphi_{1,2} && \hdots && \varphi_{1,N}\\
\varphi_{2,1} && \varphi_{2,2} && \hdots  && \varphi_{2,N}\\
\vdots && \vdots && && \vdots \\
\varphi_{2N_s,1} && \varphi_{2N_s,2} && \hdots && \varphi_{2N_s,N}  
\end{pmatrix},
\end{equation}
with $N=N_\uparrow+N_\downarrow$.
In more compact notation,
\begin{equation}
\Phi=
\begin{pmatrix}
\Phi^{\uparrow\uparrow} && \Phi^{\uparrow\downarrow} \\
\Phi^{\downarrow\uparrow} &&  \Phi^{\downarrow\downarrow}
\end{pmatrix},
\end{equation}
where in general the $\Phi^{\sigma\sigma^\prime}$ are linear combinations of  $\Phi^\uparrow$ and $\Phi^\downarrow$, 
corresponding to Eq.~\ref{eq:phi_sigma}.  In the absence of SOC, $\Phi^{\sigma\sigma}$ reduces to $\Phi^{\sigma}$,
and $\Phi^{\sigma\sigma^\prime}=0$. Because the one body operator couples spin-$\uparrow$ and spin-$\downarrow$, 
its matrix representation, $\hat{B}(\mathbf{x})$, is now $2N_s\times 2N_s$,
\begin{equation}
\begin{pmatrix}
 \mathbb{B}^\uparrow && \mathbb{B}_{SOC}^\dagger \\[1.5mm]
 \mathbb{B}_{SOC} &&  \mathbb{B}^\downarrow
\end{pmatrix}.
\end{equation}
Note that $\mathbb{B}_{SOC}$ is zero in the absence of SOC. The ensemble of random walkers,
now in the form of $2N_s\times N$ Slater determinants, is propagated and sampled until
the procedure converges to a stochastic representation of the many-body ground-state. 

Observables are computed using the Green's function between two Slater determinants, which can be measured by,
\begin{equation}
 \mathbb{G}^{\sigma \sigma^\prime}_{ij}= \frac{\langle \phi_l |c_{i\sigma}^\dagger c_{j\sigma^\prime}| \phi_r\rangle}{\langle \phi_l | \phi_r\rangle}.
\end{equation}
Each $\mathbb{G}^{\sigma \sigma^\prime}$ is a $N_s \times N_s$ matrix, obtained according to,
\begin{equation}
\begin{pmatrix}
\mathbb{G}^{\uparrow \uparrow} && \mathbb{G}^{\uparrow \downarrow} \\
\mathbb{G}^{\downarrow \uparrow} && \mathbb{G}^{\downarrow \downarrow} 
\end{pmatrix} = [ \phi_r ( \phi_l^\dagger \phi_r )^{-1} \phi_l^\dagger ]^{T}.
\label{eq:GreensFunc}
\end{equation}
As with the random walkers and one-body propagators, the application of Wick's theorem must
be modified due to the inclusion of SOC. 
Spin-flip terms in the Wick expansion,  
which make no contribution before and can be  
neglected,
must now be included.

We take as an example the calculation of double occupancy,
\begin{equation}
D_i = \frac{\langle \phi_l |n_{i\uparrow} n_{i\downarrow}| \phi_r\rangle}{\langle \phi_l | \phi_r\rangle}.
\end{equation}
Without SOC, this observable has the form, 
\begin{equation}
D_i = \mathbb{G}^{\uparrow \uparrow}_{ii} \mathbb{G}^{\downarrow \downarrow}_{ii}.
\end{equation}
In the presence of SOC it takes the expanded form, 
\begin{equation}
 D_i = \mathbb{G}^{\uparrow \uparrow}_{ii} \mathbb{G}^{\downarrow \downarrow}_{ii} - \mathbb{G}^{\uparrow \downarrow}_{ii} \mathbb{G}^{\downarrow \uparrow}_{ii}.
\end{equation}
We
note that Hamiltonians which include
pairing terms can be treated with an additional generalization,
where
the walker is 
extend to Hartree-Fock-Bogoliubov space \cite{HFB_paper}.

\section{Results}
\label{sec:results}
Having provided a brief description of the AFQMC method, we present in the
following sections an illustration of its capabilities by highlighting a small set of the many 
observables that the method makes possible to measure to high accuracy. 
We discuss results for ultracold atoms and the repulsive electronic model in two separate
subsections.

\subsection{Fermi atomic gas with synthetic SOC}
\label{ssec:Fermigas}

We begin with
a survey of recent results on ultracold atoms, where we will focus on dilute Fermi gas systems in 
2D. Further results
can be found in \cite{2DFG_SOC_AFQMC},
including results for optical lattices \cite{2DFG_RASHBASOC_OPLATT_AFQMC}.
As mentioned in Sec.~\ref{sec:Hamiltonian}, the Hamiltonian has $U/t<0$ here. The calculations
are performed at the dilute limit, with $N\ll N_s=L^2$. In order to map the results to the continuum 
system of Fermi atomic gases studied in experiment,
we first introduce an overall energy scaling defined by the ground-state energy per particle of the 
corresponding non-interacting Fermi gas,  $E_{FG}=\pi n$, with $n=N/L^2$ being the number density for 
the 2D lattice. 
The interaction strength $U$ is uniquely defined \cite{WernerCastin,2DFG_AFQMC} by $\log(k_F a)$.  
In order to compare physically equivalent systems we introduce two dimensionless parameters:
\begin{align}
\alpha = \frac{\lambda^2}{E_{FG}}; \quad
\eta = \frac{\varepsilon_B}{E_{FG}},
\end{align}
that specify the strengths of the SOC and interaction, respectively.
In the above, $\varepsilon_B$ is the two-body binding energy at $\lambda=0$ and 
$\varepsilon_B/E_{FG}$
is directly related to $k_F a$ 
\cite{2DFG_AFQMC}. The strong SOC regime, characterized by occupation of only the $\varepsilon_\mathbf{k}^-$ 
helicity band,
 occurs for $\alpha>4$. In the weak SOC regime ($\alpha<4$), both $\varepsilon_\mathbf{k}^-$ and 
$\varepsilon_\mathbf{k}^+$ are occupied. There is a smooth transition between the two regimes at $\alpha = 4$.
Small $\eta$ corresponds to the BCS limit of the BCS-BEC crossover, where the physics is best described
in terms of weakly interacting Cooper pairs, whereas large $\eta$ corresponds to the BEC limit,
in which strongly interacting fermions are bound into tight pairs resembling bosons.

\begin{figure}
  \includegraphics[width=\columnwidth]{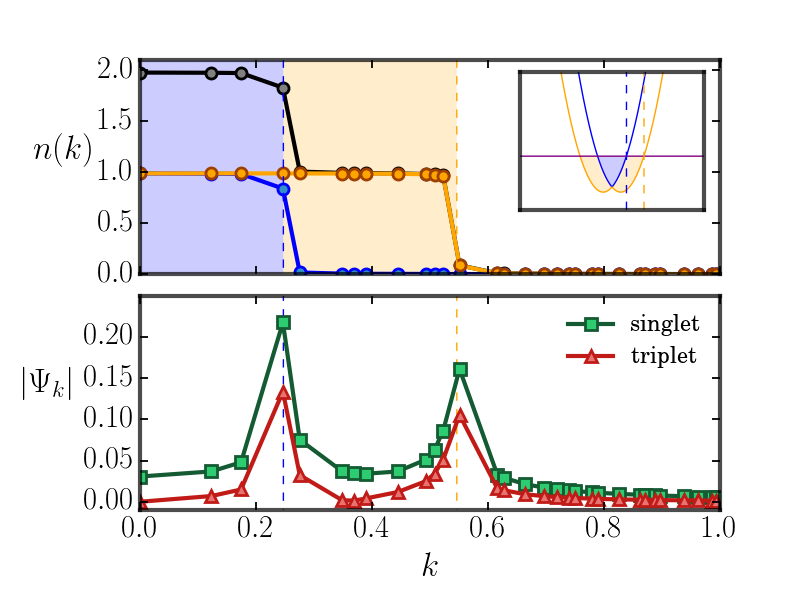}
  \caption{Momentum distributions and pairing wavefunctions at $\alpha=1.0$, $\eta=0.001$. The top row 
  shows the total momentum distribution in black, and the occupation of $\varepsilon^+_\mathbf{k}$ and 
  $\varepsilon^-_\mathbf{k}$ in blue and orange respectively. The non-interacting dispersion is shown in the inset, 
  with the corresponding regions of occupation and the Fermi surfaces indicated by the shading and dashed 
  lines in the main plot. The bottom row plots the singlet and triplet components of the pair wavefunction. 
  The system has  74 particles, using a $51\times51$ periodic supercell.}
  \label{fig:momentum_plus_pairing_b0.001}
\end{figure}

\begin{figure}
  \includegraphics[width=\columnwidth]{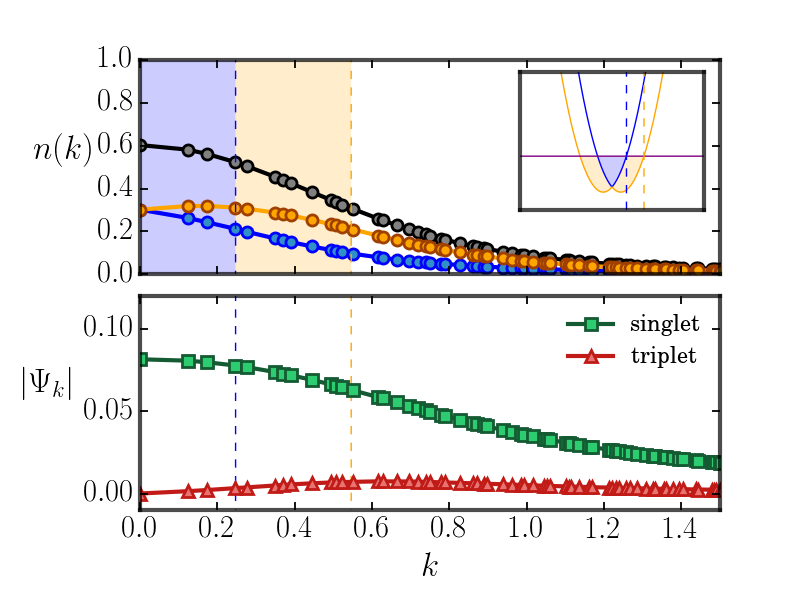}
  \caption{Momentum distributions and pairing wavefunctions at $\alpha=1.0$, $\eta=10.0$.
  The remaining system parameters are the same as in Fig.~\ref{fig:momentum_plus_pairing_b0.001}.}
  \label{fig:momentum_plus_pairing_b10.0}
\end{figure}

Figures \ref{fig:momentum_plus_pairing_b0.001} and \ref{fig:momentum_plus_pairing_b10.0} 
plot two examples of the computed momentum distribution and pairing wavefunctions. The first figure corresponds
to a weakly interacting system, with $\eta=0.001$, and the second to a strongly interacting system, 
with $\eta = 10.0$. Both systems are in the modest 
SOC regime, with $\alpha=1.0$. Large lattice sizes are treated here, so that the residual 
discretization error and finite-size (number of particles) error are small.
The results provide quantitative 
information for the continuum bulk 2D Fermi gas.

The momentum distribution 
in the weakly interacting system differs only slightly from the non-interacting case, whose regions of occupation 
and Fermi surfaces (one for each helicity band) are indicated by the shading and vertical dashed lines. 
At large interaction strength the momentum distribution reveals that there is occupation of higher momentum 
states that are well beyond the non-interacting region. In addition to becoming considerably broader
as a function of interaction strength, the distribution also becomes smoother, showing no
evidence of the cusps present at small interaction strengths.

AFQMC also provides access, at a quantitative level, to the rich pairing structure induced by 
the presence of SOC.
The bottom row of figures \ref{fig:momentum_plus_pairing_b0.001} and
\ref{fig:momentum_plus_pairing_b10.0} present two examples of the pairing wavefunction. SOC 
mixes the singlet and triplet channels, so the pairing wavefunction has both singlet and triplet components. At 
small interaction strengths the pairing wavefunction has strong peaks indicating that pairing is concentrated near
the Fermi surfaces. Also evident is the fact that the singlet and triplet components of the pairing wavefunction
have similar amplitude. This is no longer the case for strongly interacting systems, which favor singlet pairing over 
triplet pairing
because of the enhanced on-site attraction, as reflected by the relative amplitude of the two components of the pairing wavefunction. As a
reflection of the behavior of the momentum distribution, the well-localized peaks of the pairing wavefunction also 
become broad and smooth at large interaction strength, suggesting that pairing occurs across a wide range of momenta.

\begin{figure}
  \includegraphics[width=\columnwidth]{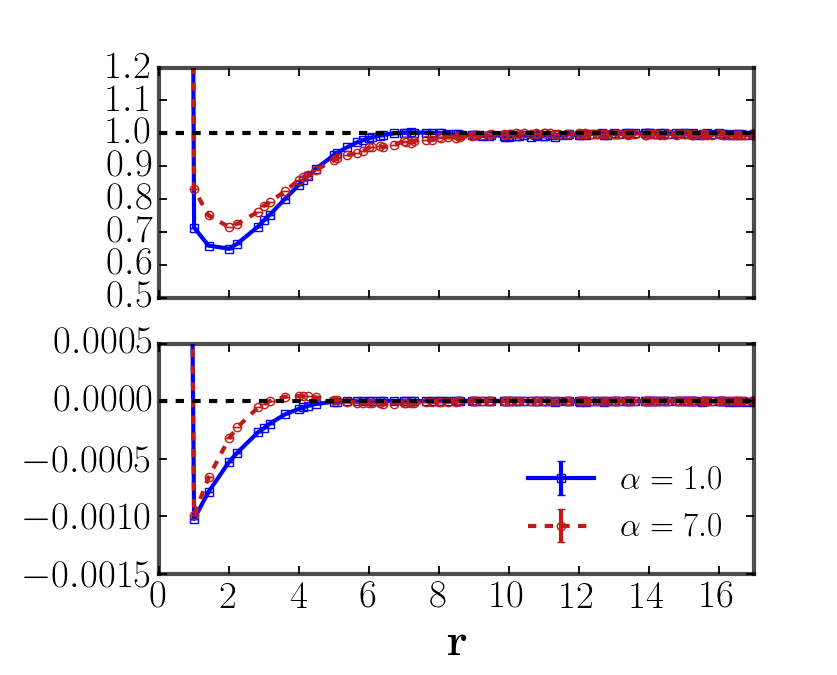}
  \caption{Charge-charge and spin-spin correlations vs.~$\alpha$ for
  interaction strength, $\eta=0.001$. The top panel plots $\langle n_0 n_r \rangle/n^2$ 
  and the bottom panel plots $\langle\vec{S}_0\cdot\vec{S}_r\rangle$. The system
  has  58 particles for $\alpha=1.0$ and 56 particles
  for $\alpha=7.0$, using 
  a $35\times35$ periodic supercell.}
  \label{fig:charge-charge_spin-spin_b0.001}
\end{figure}

\begin{figure}
  \includegraphics[width=\columnwidth]{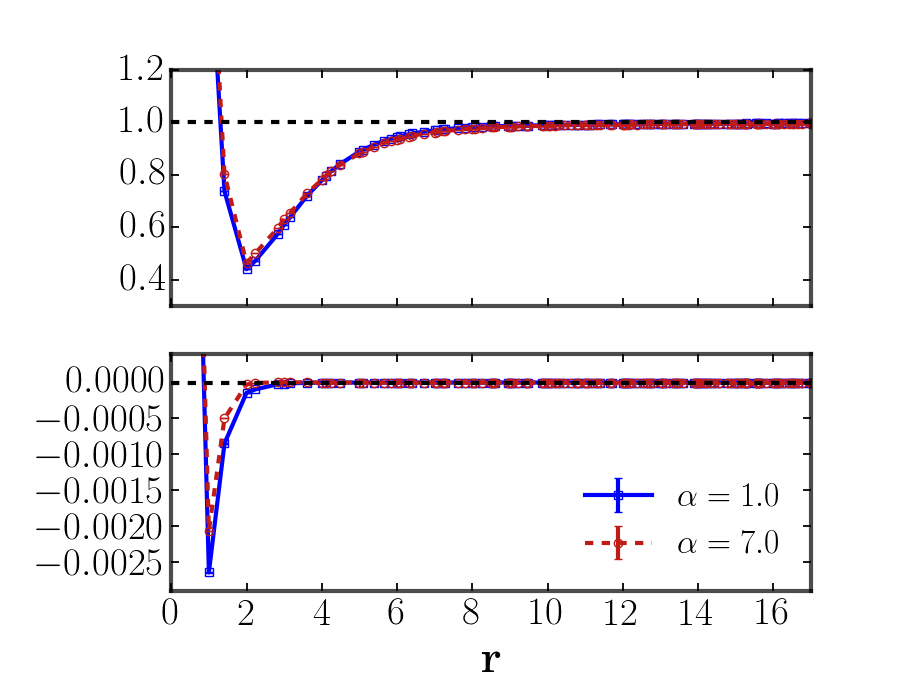}
  \caption{Charge-charge and spin-spin correlations vs.~$\alpha$ for
  interaction strength, $\eta=10$. The top panel plots $\langle n_0 n_r \rangle/n^2$ 
  and the bottom panel plots $\langle\vec{S}_0\cdot\vec{S}_r\rangle$. The remaining
  system parameters are identical to those in Fig.~\ref{fig:charge-charge_spin-spin_b0.001}.}
  \label{fig:charge-charge_spin-spin_b10}
\end{figure}

We can also reliably measure many other observables, including charge-charge and spin-spin correlation 
functions. Figures~\ref{fig:charge-charge_spin-spin_b0.001} and
\ref{fig:charge-charge_spin-spin_b10} illustrate the effects of SOC on the charge-charge and spin-spin 
correlations at two different values of interaction strength. At weak interaction
(Fig.~\ref{fig:charge-charge_spin-spin_b0.001}) and weak SOC, the charge-charge correlation
exhibits a small amplitude oscillation about the square of the average density. This oscillation is
suppressed as SOC strength is increased. The spin-spin correlation shows an opposite effect,
with a small oscillation about zero at large SOC strength that is not present for weak SOC.
With increased interaction strength (Fig.~\ref{fig:charge-charge_spin-spin_b10}), both correlation
functions become shorter ranged with larger amplitude, which is evident in the steeper slopes
of each as they approach their asymptotic values.

\subsection{Towards real materials --- the repulsive Hubbard model with SOC}
\label{ssec:Hubbard}

\begin{figure}
  \includegraphics[width=\columnwidth]{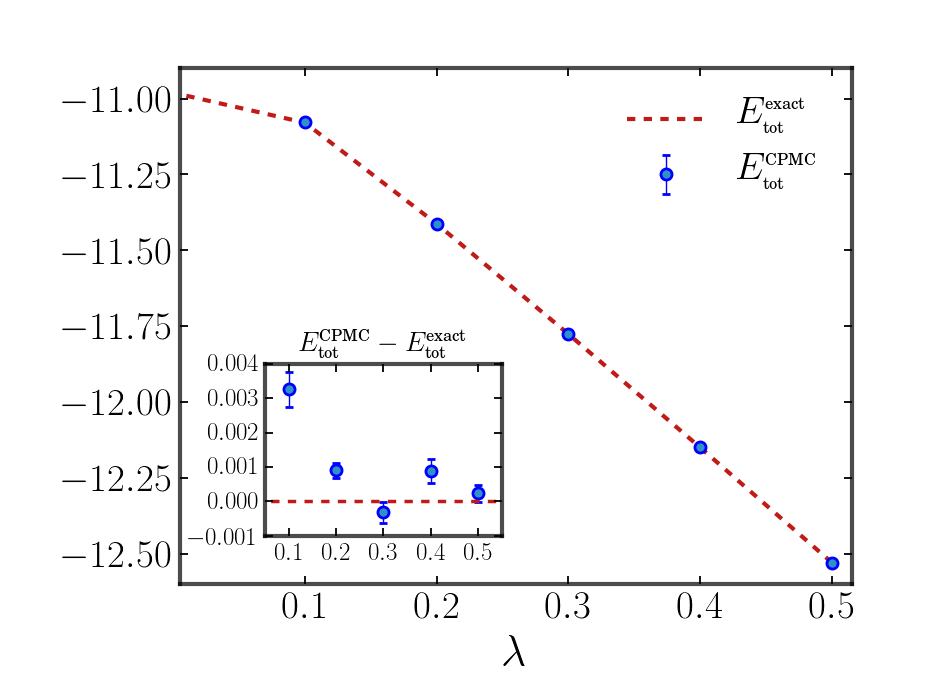}
  \caption{Calculated total energy as a function of SOC strength $\lambda$ at interaction strength $U/t=4$.
  The inset shows the discrepancy compared to exact results.}
  \label{fig:en_vs_lbda}
\end{figure}

\begin{figure}
  \includegraphics[width=\columnwidth]{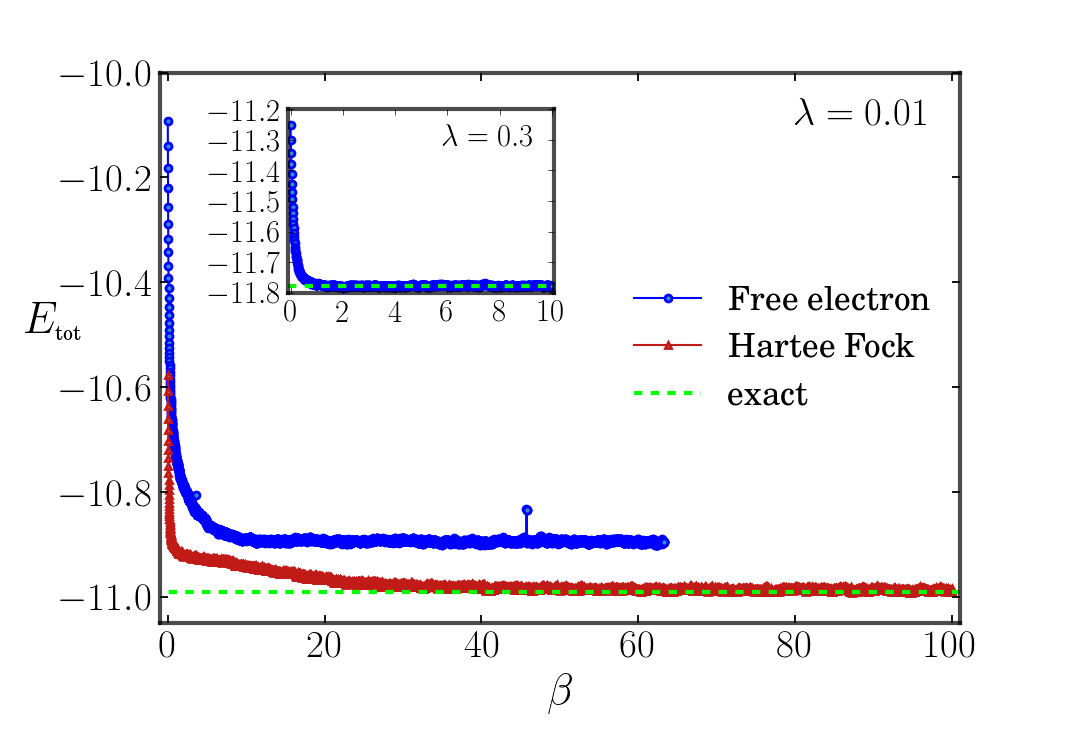}
  \caption{Energy as a function of imaginary time $\beta$ at $\lambda=0.01$ and $U/t=4$.
  	The blue curve corresponds to a simulation with a free-electron wavefunction used as the trial wavefunction,
	and the red curve corresponds to a simulation with a generalized Hartree-Fock wavefunction as the trial wavefunction.
	The inset shows the energy versus projection time $\beta$ at $\lambda=0.3$ and $U=4$, with a free-electron trial
	wavefunction.}
  \label{fig:en_vs_beta_lbda0.01_lbda_0.3}
\end{figure}

As a demonstration of the applicability of this approach to electronic systems, we present a study of the
repulsive Hubbard model with Rashba SOC. We first examine the energetics, which
highlight the capability of CP-AFQMC to achieve high-accuracy results for systems with a sign 
problem. To benchmark our method we focus on a $2\times 4$ supercell with $N=4$ electrons, for which we perform 
exact diagonalization calculations for comparison. 

\begin{figure}
  \includegraphics[width=\columnwidth]{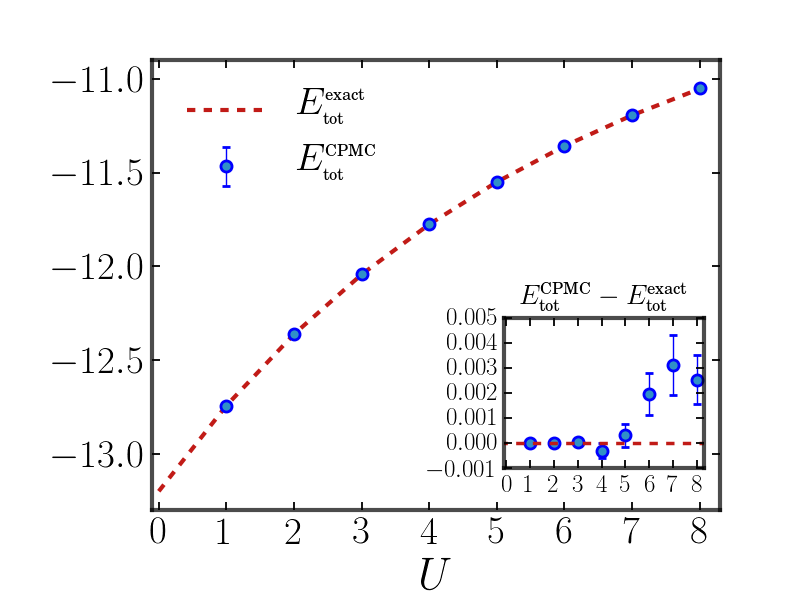}
  \caption{Total energy as a function of interaction strength $U/t$ at SOC strength $\lambda=0.3$.
  The inset shows the difference between the energy obtained from CP-AFQMC and the exact energy.}
  \label{fig:en_vs_U}
\end{figure}

Plotted in Fig.~\ref{fig:en_vs_lbda} is the total energy as a function of SOC
strength at an interaction strength $U/t=4$,
in comparison with exact results. 
In this set of simulations the free-electron wavefunction was
used as the trial wavefunction $|\Psi_T\rangle$  (see Eq.~(\ref{eq:constraint})).
Very good agreement is seen with exact results.
The slightly larger discrepancy at small SOC strength can be improved with an improved trial wavefunction,
as illustrated in Fig.~\ref{fig:en_vs_beta_lbda0.01_lbda_0.3}. In this case a generalized Hartree-Fock
solution was used as $|\Psi_T\rangle$
and the CP-AFQMC energy shows improved convergence
to the exact energy. 
The behavior of the algorithm as a function of interaction strength is investigated in Fig.~\ref{fig:en_vs_U}, which shows the total energy vs.~$U/t$ at a fixed SOC strength of $\lambda=0.3$. 
Nearly exact results are obtained for up to intermediate interaction. Even in the strong interaction limit,
the relative error is a small fraction of  a percent, which is well within the accuracy of AFQMC seen 
in Hubbard-like models \cite{MingpuHubbardBenchmark, Hao-symmetry} or in molecules and solids 
\cite{AFQMC_backprop_molecules,HydrogenBenchmark,AFQMC_downfolding}.

\begin{figure}
  \includegraphics[width=\columnwidth]{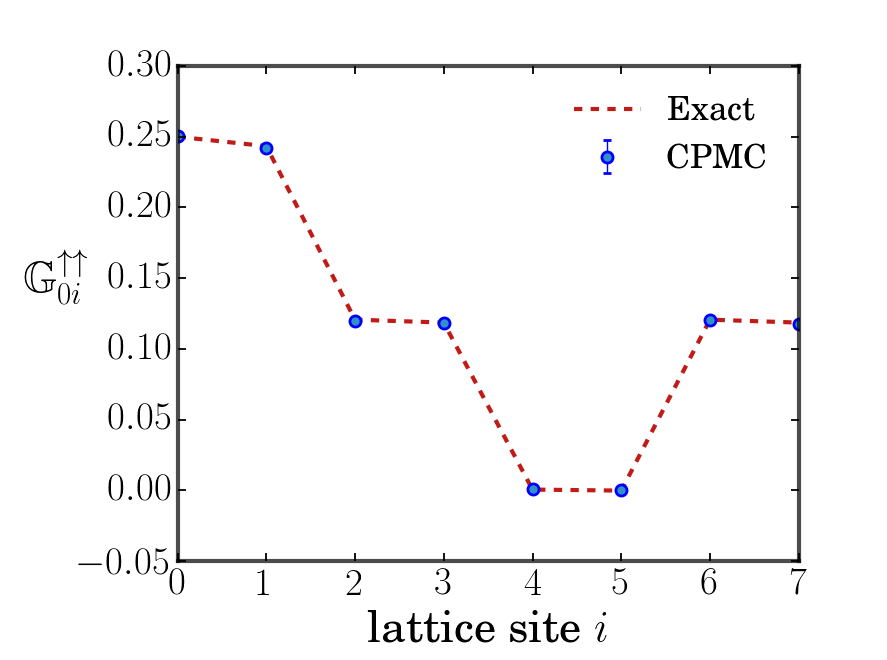}
  \caption{Green's function at interaction strength $U/t=4$ and SOC strength $\lambda=0.3$.
  The first index is fixed at lattice site 0 and the second index runs over all sites.}
  \label{fig:Guu}
\end{figure}

In addition to ground-state energies, high-accuracy measurements of other observables,
including charge, spin and pairing properties, are possible with CP-AFQMC. As indicated in the procedure 
outlined above, observables can be obtained from the Green's function, which is calculated according to 
(\ref{eq:GreensFunc}). We show an example of the Green's function in Fig.~\ref{fig:Guu}. 
Comparison 
of the CP-AFQMC result with the exact result establishes the high degree of accuracy attainable in CP-AFQMC 
measurements of observables.   

\section{Summary and Outlook}
\label{sec:summary}

Spin-orbit coupling is an essential ingredient of many fascinating phenomena across a wide
range of physical contexts. From spintronics to topological insulators to quantum information, a precise
characterization of the behavior of systems subject to SOC will not only improve the fundamental
understanding of these phenomena and phases, it will also inspire and guide the fabrication 
of new devices.

The results presented here show the applicability and the accuracy of the AFQMC method, which
provides precision, and in some cases, such as the attractive 2D Fermi gas, numerically exact treatments 
of strongly interacting many-body systems. As we have demonstrated, AFQMC is well suited to study lattice models 
with either attractive or repulsive interactions. More realistic treatments of materials can be achieved by replacing the 
on-site Hubbard interaction with generalized (4-index) two-body matrix elements. Interactions of this generalized form 
can be incorporated using any single-particle basis, such as plane waves or Gaussians, but require a new decomposition
and Hubbard-Stratonovich transformation, which introduces complex auxiliary fields and consequently
a phase problem. This problem can be systematically controlled using a generalization of the CP approach, 
the phaseless approximation \cite{PRL-phaseless}, 
which can achieve similar accuracy to what is illustrated above, in realistic material simulations \cite{HydrogenBenchmark,
AFQMC_downfolding}. Our discussion of the treatment of SOC carries through straightforwardly,
using GHF-type walkers.  As we have demonstrated, this approach enables the treatment of SOC with no degradation of the 
exquisite accuracy achievable in electronic structure calculations with AFQMC.

As the experimental and theoretical landscape continues to expand, there will be many opportunities
and challenges in the study of SOC and strong interaction in many-body systems. These challenges,
including characterizing the effect of interaction on topological phases \cite{PhysRevB.89.195124,PhysRevB.90.245120,
annurev-conmatphys-SPT}, will require complementary theoretical, numerical, and experimental progress to generate
new understanding. Given its unique ability to treat strongly-correlated many-body systems with high precision, 
AFQMC will play a vital role in the combined effort.  
\\
\\
\textit{Acknowledgments} This research was supported by NSF (grant no.~DMR-1409510), and the Simons Foundation. Computing was
carried out at the Extreme Science and Engineering Discovery
Environment (XSEDE), which is supported by
NSF grant number ACI-1053575, and the computational
facilities at the College of William and Mary. The Flatiron Institute is supported by the Simons Foundation.

\bibliography{AFQMC_OPLATT}

\end{document}